\documentclass[5p, twocolumn, times]{elsarticle}

\usepackage{graphicx}
\usepackage{amsmath}  
\usepackage{tabularx} 		

\begin{document}


\begin{frontmatter}

	\title{Uniformity Studies of Scintillator Tiles directly coupled to SiPMs for Imaging Calorimetry}
	
	\author[add1,add2]{F.~Simon}
	\author[add1,add2]{C.~Soldner\corref{cor}}
	\ead{soldner@mpp.mpg.de}
	
	\cortext[cor]{Corresponding author}
	
	\address[add1]{Max-Planck-Institut f\"ur Physik, M\"unchen, Germany}
	\address[add2]{Excellence Cluster Universe, Technische Universit\"at  M\"unchen, Garching, Germany}

	\begin{abstract}
	
We present a novel geometry of scintillator tiles developed for fiberless
coupling to silicon photomultipliers (SiPMs) for applications in highly
granular calorimeters. A high degree of uniformity of the tile response over the full active area was
achieved by a drilled slit at the coupling position of the photon sensor with
$2\,\text{mm}$, $4\,\text{mm} $ and $5.5\,\text{mm} $ in height, width and depth.
Detailed measurements of the response to penetrating electrons were performed for
tiles with a lateral size of $3\times3\,\text{cm}^2 $ and thicknesses of $
5\,\text{mm} $ and $ 3\,\text{mm}$.
	
	\end{abstract}
	

\end{frontmatter}

\section{Introduction}

The physics at a future high-energy linear $e^+e^-$ collider imposes stringent
requirements on the detector performance. The detailed study of physics beyond
the Standard Model as well as precision measurements of Standard Model processes
demand excellent reconstruction of multi-jet final states with missing transverse
energy, well beyond the performance of present high energy particle physics
detector systems. This improved jet energy resolution is achievable with particle
flow algorithms (PFA) \cite{Brient:2002gh, Morgunov:2002pe, Thomson:2009rp}, which make optimal use of all
available information in one event by reconstructing every single particle in a
jet using the best energy measurement for each particle type. For charged
particles, the tracker information is used since this provides the highest
precision, while neutral hadrons and photons are reconstructed in the
calorimeters. This technique demands the separation of the showers of individual
particles in a jet, which in turn requires extreme granularity in the
calorimeters. The high granularity provides a detailed three dimensional image of
the energy deposits which allows the separation of showers and thus the
unambiguous assignment of calorimetric clusters to particle tracks identified in
the tracking system.\\
A first physics prototype of a highly granular analog hadron calorimeter has been
tested extensively by the CALICE collaboration \cite{CALICE:AHCAL}. This
calorimeter uses small cells of conventional plastic scintillator as active
medium, each individually read out by silicon photomultipliers (SiPMs)
\cite{Golovin:2004jt}, coupled to a wavelength shifting fiber which is embedded
in the scintillator tile as shown in Figure
\ref{fig:Analysis:WLSTile}. The fiber was necessary to match the wavelength of
the emitted photons to the sensitivity maximum of the used SiPMs in the green
spectral range. The size of the scintillator tiles has been optimized with PFA
simulations to be $3\times3\,\text{cm}^2$ \cite{Thomson:2009rp, ILC:LoI}. For a
full hadron calorimeter system this implies several million cells, requiring a
simple cell design well suited for mass production.\\
The recent development of blue-sensitive SiPMs, such as the Multi-Pixel Photon
Counter (MPPC) from Hamamatsu \cite{Hamamatsu}, opens up the possibility for
direct fiberless coupling of the photon sensor to the scintillator cell, relaxing
the mechanical tolerances on the SiPM positioning and simplifying the mass
production of cells. However, a wavelength shifting fiber embedded in the
scintillator also improves the light collection, leading to a largely uniform
response of the tile to penetrating particles over the full active area. A
strongly non-uniform tile response, which can be introduced by a direct coupling
of the photon sensor, leads to a distortion of the reconstructed energy in a
complete calorimeter, and in addition also compromises the calibration of the
detector cells based on single particle signals. It is therefore important to
design a SiPM-scintillator tile assembly that provides a spatially uniform
response, also in the case of fiberless coupling of the photon sensor.

It has been shown that a high degree of uniformity can be achieved with
specialized shaping of the scintillator tiles in the case of a direct coupling on
the large surface of the cell  \cite{Blazey:2009zz}. In the following, the
performance of directly coupled square scintillator cells, which are $
9\,\text{cm}^2 $ in area and $3$ or $ 5\,\text{mm} $ thick, with the photon
sensor coupled to one side face, are investigated in detail. The cell geometry
was modified from a simple tile shape and optimized in terms of uniformity of the
tile response, photon collection efficiency and electro-mechanical integrability.
While the studies presented here were performed in the context of a highly
granular hadronic calorimeter for experiments at a future linear $e^+e^-$
collider, they also apply to other calorimetric systems as well as to other
detectors in high energy physics using plastic scintillator technology.

\section{Test Setup}

\begin{figure}
	\centering 
	\includegraphics[width=.99\linewidth]{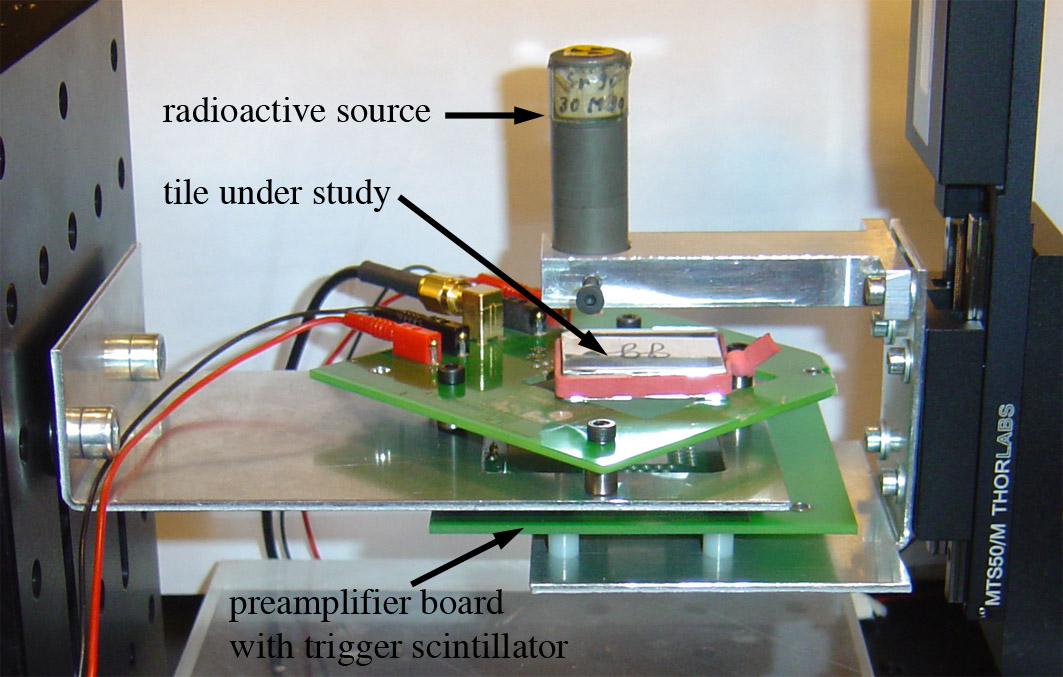}
	\caption{Photograph of the experimental setup, showing the $^{90}$Sr 
	source, the tile under study and the mechanical support structures 
	including the translation stage.}
	\label{fig:FotoSetup}
\end{figure} 

\begin{figure}
	\centering 
	\includegraphics[width=.65\linewidth]{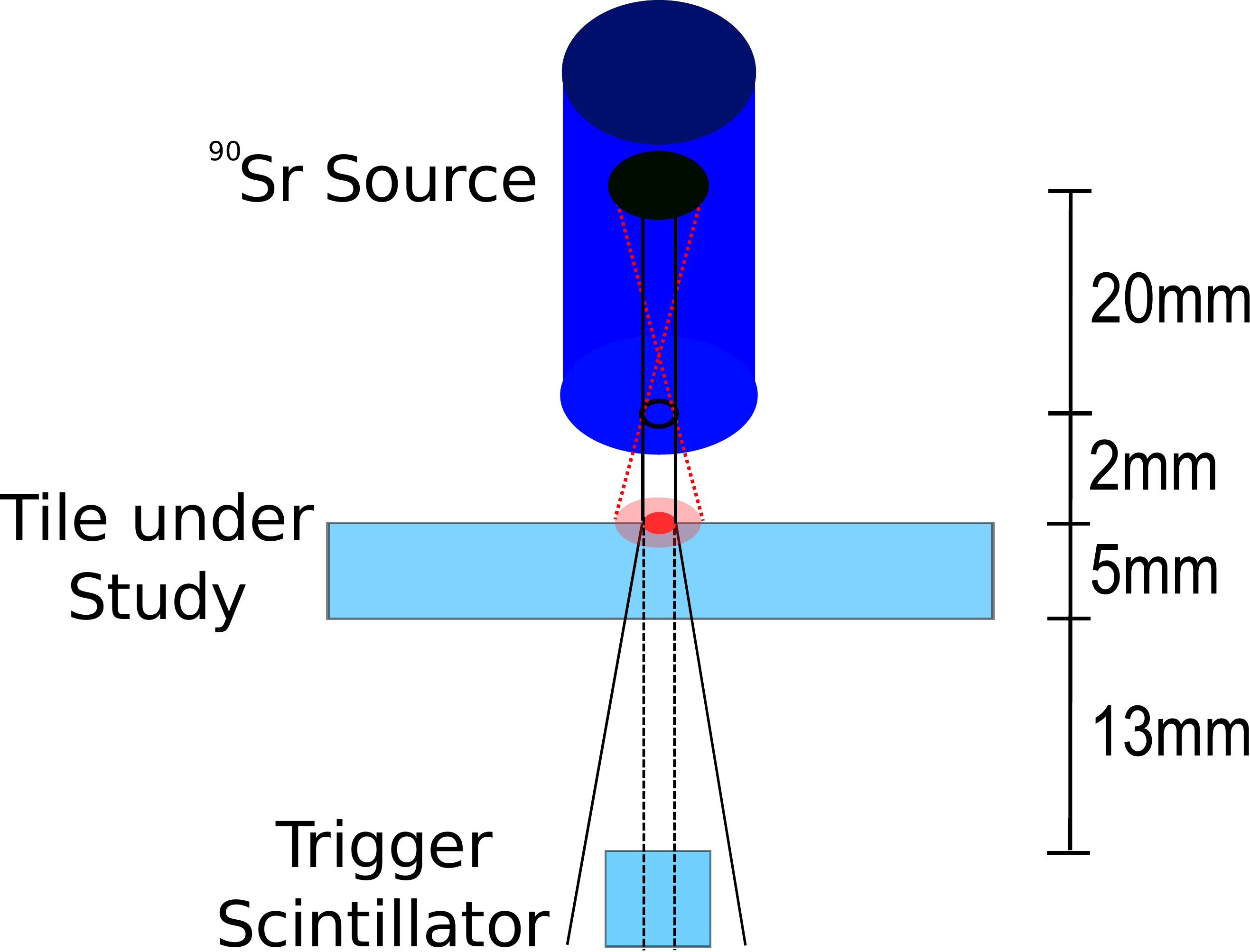}
	\caption{Schematic illustration of the experimental setup used for the 
	measurement of the uniformity of the tile response to penetrating electrons 
	(not to scale).}
	\label{fig:ExpSetup}
\end{figure}

The spatial dependence of the response of scintillator tiles with directly
coupled SiPMs to penetrating charged particles was studied with a $^{90}$Sr
source, which emits electrons with an end point energy of 2.27 MeV,  yielding a
sufficient fraction of particles that are capable of penetrating the scintillator
tiles under study completely. The source emitted electrons through a circular
collimator opening with a diameter of $ 1\,\text{mm} $. A cubic scintillator with an edge
length of $ 5\,\text{mm} $ was positioned in $13\,\text{mm} $ distance
underneath the tile under study as an additional trigger detector. The source and
the trigger scintillator were mounted on a high precision translation stage,
allowing to scan the source across the surface of the tile under study. A
photograph of the experimental setup, including the support structures and the
translation stage, is shown in Figure \ref{fig:FotoSetup}, while Figure
\ref{fig:ExpSetup} shows a schematic of the arrangement of the source, the tile
under study and of the trigger scintillator.

Both the trigger scintillator and the different $3\times3\,\text{cm}^2$ tiles
that were studied here were  fabricated from 5 mm thick plates of Saint-Gobain
BC-420 premium plastic scintillator \cite{Scinti:Datasheet}. The scintillator was
enclosed with an aluminized reflective foil \cite{Tile:3MFoil} on all sides,
leaving only a small opening for the coupling of the photon sensor in the case of
a simple direct coupling, and completely enclosing the tile in the case of SiPM
integration as discussed below.

Light detection was performed by Hamamatsu MPPCs with an active area of 1\,mm$^2$
and a pixel size of $25\times25\,\mu\text{m}^2 $, resulting in 1600 pixels per
sensor. For the read out of the trigger scintillator, an MPPC in a ceramic
package (type S10362-11-025C) was mounted underneath the plastic cube. The tiles
under study were read out with an MPPC in a clear plastic package  with a size of
$ 3\, \times 4\, \text{mm}^2$ with a thickness of 1.5\,mm, a prototype of the now
available SMD-mount type S10362-11-025P.  In all presented measurements, the
MPPC-25P was coupled directly to the center of one side face of the tile.

The MPPCs were mounted on custom designed preamplifier boards with a fixed
amplification factor of 8.9. A 4-wire temperature sensor (thermistor) was placed
in thermal contact with the MPPC used to read out the tile under study to monitor
changes in the temperature, allowing an offline correction for the temperature
dependence of the device response. The signal readout was performed with a $
4\,\text{GHz} $, 20 GSamples/s oscilloscope with an acquisition time window of $
50\,\text{ns} $ per recorded event. The readout was triggered by requiring a
coincidence of signals above the equivalent of two detected photons in the tile under study and in the
trigger cube. The requirement of a coincidence in both the tile and the trigger
scintillator ensured the selection of signals from electrons that fully traversed
the tile under study. Due to the energy spectrum of the electrons emitted from
the $^{90}$Sr source, most of the produced electrons are absorbed fully in the
tile, making this coincidence selection mandatory to obtain a good sample of
penetrating particles.

\begin{figure}
 	  \centering
 	  \includegraphics[width=.9\linewidth]{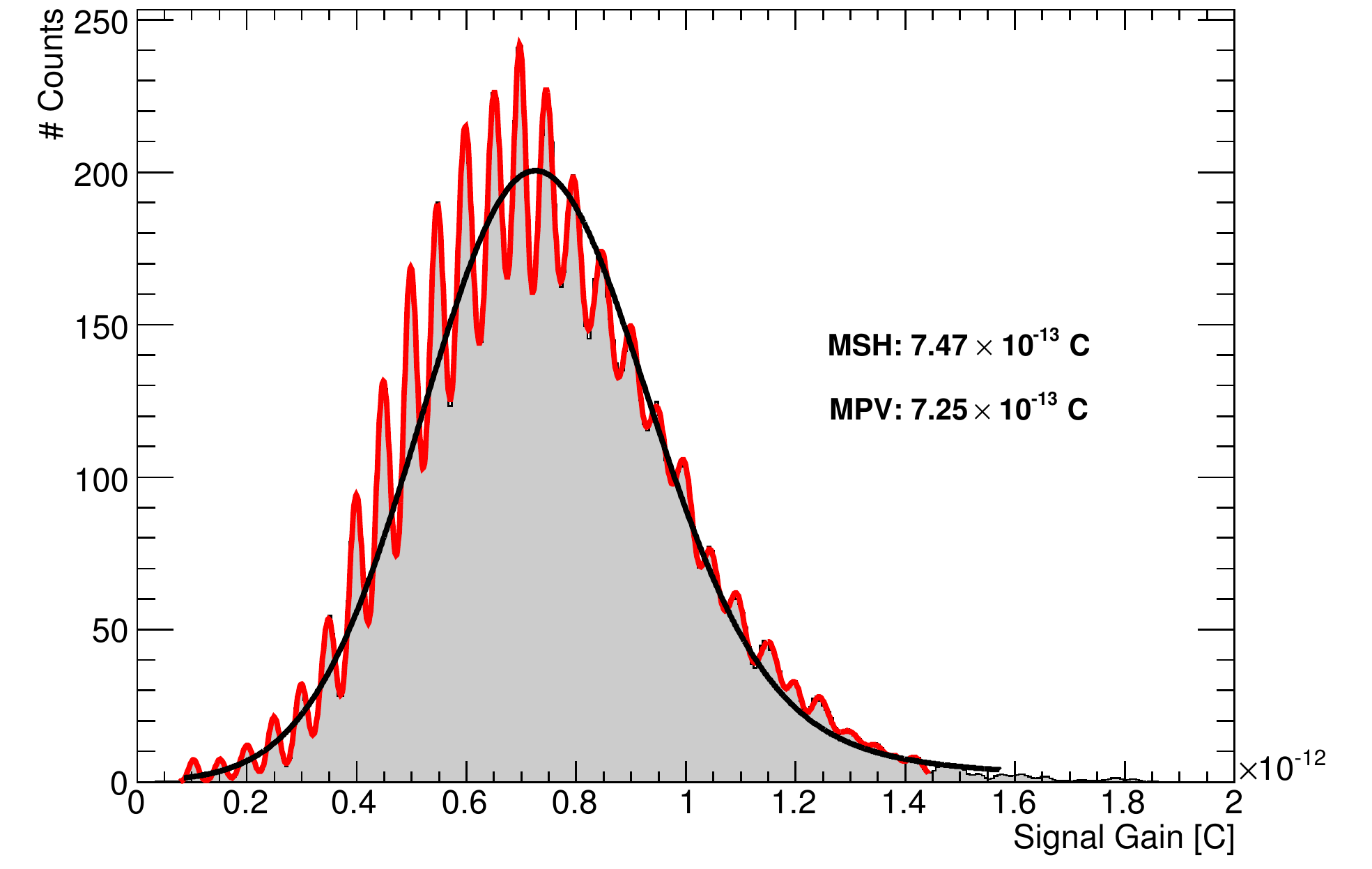}
 	  \caption{Distribution of recorded charge for a high statistics run at a 
 	  representative position of a modified tile (as discussed in the text), clearly 
 	  showing the individual peaks of a given number of detected photons. The 
 	  most probable value (MPV) of the distribution is extracted from the shown
 	  fit of a Landau convoluted with a Gaussian. The mean signal height ({\it MSH}) 
 	  is given by the mean of the distribution. Also shown is a multi-Gauss fit used to
 	  extract the peak-to-peak distance to determine the gain of the SiPM.}
 	  \label{fig:Exp:PulseDist}
 \end{figure} 

The total charge of each electron signal was determined in an offline analysis
procedure from the integral of the recorded wave form. The distribution of the
recorded charge for 20\,000 events , with the radioactive
source located above the center of the tile under study, is shown in Figure
\ref{fig:Exp:PulseDist}, together with a fit of a Landau function convolved with
a Gaussian. The distribution shows well-separated peaks corresponding to a
certain number of fired pixels within the MPPC. A multi-gauss fit
was applied to extract the peak-to-peak charge difference, which was used to determine  the gain
of the SiPM device. The measured gain was used to rescale the
obtained signal charge in units of photon equivalents. The
overall spectrum is a convolution of the distribution of the
energy deposition in the tile by the penetrating electrons, of the photon
collection statistics and of electronic noise and signal integration
uncertainties. The latter two can be seen from the width of the individual photon
peaks. The recorded energy spectrum does not exhibit the pronounced long tail to
high energy deposits  characteristic for minimum ionizing particles in thin
absorbers. This is due to the low energy of the electrons used in the
measurement. The mean energy loss of a minimum ionizing particle in the 5 mm
thick scintillator tile is around 1 MeV, thus the requirement that the particle
penetrates the tile and generates a sizeable signal in the trigger scintillator
limits the amount of energy deposited in the tile under study, essentially
cutting off the high energy tail of the energy loss distribution.

\section{Measurement of the Uniformity of Response}

The cell uniformity, which is the response of the scintillator tile with respect
to the particle impact position, was determined through a lateral scan of the
radioactive source across the surface of the cell under study. The source, and
with it the trigger scintillator, were moved with a step size of $0.5\,\text{mm}
$ in a rectangular pattern across the tile surface, covering the full active area
including the edge regions. The source was steered to
$61\times61$ positions on the tile, recording 500 electron
signals and the current temperature at each point. Before every
tile scan, the center of the source casing and therefore the center of the
distribution of emitted electrons was aligned to the center of the tile under
study. Consequently, the source was aligned to the tile edges for the outermost
on-tile positions.

The position dependent mean signal height ({\it MSH}) was extracted from the
signal distributions recorded at each point.  Since the statistics of 500 events
is not sufficient to perform a full fit of the distribution as shown in Figure
\ref{fig:Exp:PulseDist}, the mean of all recorded amplitudes is used to evaluate
the response at each position.

\begin{figure*}
	  \centering
	  \includegraphics[width=0.85\linewidth]{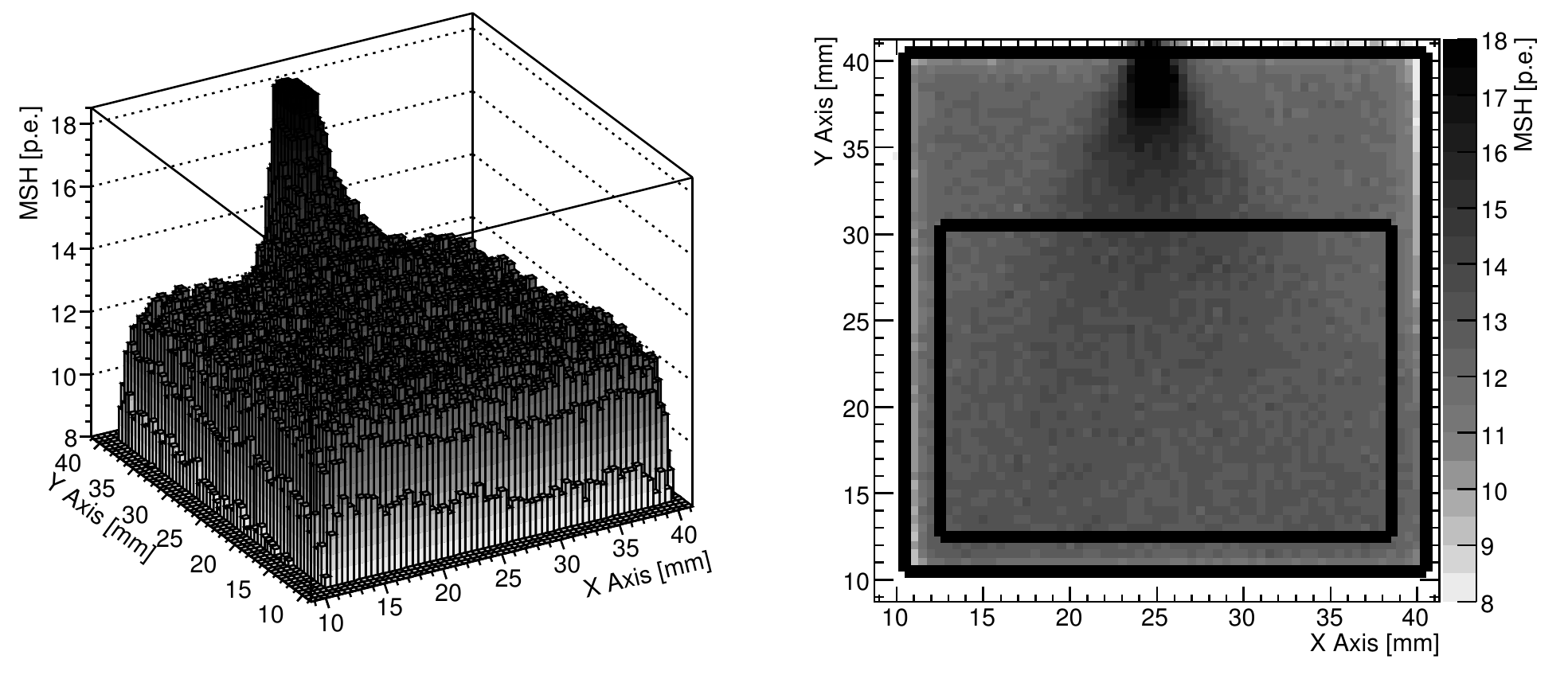} 
	  \caption{Direct SiPM coupling to an unmodified tile: Mean signal height of 
	  minimum ionizing electrons in dependence of the penetration position (3D and 2D 
	  view). The outer black square in the 2D plot identifies the tile position, the 
	  inner black rectangle shows the region over which an average is performed to 
	  obtain the overall mean of the signal height, as discussed in the text.}
	  \label{fig:Analysis:UnmodTile}
\end{figure*}

Figure \ref{fig:Analysis:UnmodTile} shows the spatial distribution of measured
{\it MSH} for a simple unmodified tile with the dimensions $
3\times3\times0.5\,\text{cm}^3$, flat polished surfaces and the MPPC-25P coupled
to one flat side face of the tile. This version of the scintillator tile
corresponds to the simplest possibility for a direct coupling of the photon
sensor on the side of the tile, and is referred to in the following as `Simple
Tile 5' or `ST5', where the ``5'' denotes the thickness of 5 mm. A pronounced
non-uniformity of the response as a function of the position of the particle
incidence is clearly apparent, with a significantly increased signal amplitude
close to the coupling position of the photon sensor.

Since the response of SiPMs shows a strong temperature dependence, the data were
corrected for the ambient temperature with an offline correction procedure. The
temperature of the SiPMs varied by about $ \pm2\,^{\circ}\text{C} $ during a
standard tile scan due to  day-night variations in the laboratory. The dependence
of the amplitude {\it A}, which is the most probable signal height extracted from
the signal distribution of the SiPM-Tile entity, on temperature was obtained for
each individual SiPM device in a separate run by recording data with the source
at a representative position on the tile over an extended period with changing
temperatures. In the temperature range of interest here the relative temperature
dependence of the amplitude $\frac{1}{A}\frac{dA}{dT}$ is approximately constant,
and was measured to be -4.4\%/K for the MPPC-25P. The {\it MSH} was then
corrected by a scale factor S calculated for each position:
\begin{equation*}
		\text{\it MSH}_{Corr}(x,y) = \text{\it MSH}_{Uncorr}(x,y)\cdot S , 
\end{equation*}
where 
\begin{equation*}
S = \left( 1 + \frac{1}{A} \frac{dA}{dT} \cdot \Delta T\right)^{-1}
\end{equation*}
with $\Delta T$ giving the difference of the temperature at the point of
measurement to the mean temperature over the full measurement period.

\section{Data Analysis}

A method was developed to quantify the response and uniformity of a tile and
enable a comparison of the performance of different tile geometries. The response
of the tile ST5, with a directly coupled photon sensor without any modifications
to the scintillator is shown in Figure \ref{fig:Analysis:UnmodTile}. It is
characterized by a steep overshoot at the SiPM coupling position, a sharp drop at
the tile edges and a quasi-constant response over most of the remaining active
area. To characterize the mean response of a tile to penetrating minimum ionizing
particles without strong influence from the non-uniformities due to the photon
sensor coupling, the overall mean signal height ({\it OMSH}) for a given tile is
defined as the mean value of all measurement points outside the extreme regions
of the distribution, taken over the inner rectangle shown in Figure
\ref{fig:Analysis:UnmodTile}. The exact location of the tile within the scanned
area can also be clearly identified by the signal drop at the tile edges and is
illustrated by a black square with an edge length of $3\,\text{cm} $.

The mean signal amplitude for a penetrating minimum ionizing particle is a
crucial performance parameter of the SiPM-scintillator tile entity. It is a
compromise between a high signal-to-noise ratio for single
particles and a large dynamic range, limited by
the finite number of pixels on the photon sensor. The latter
requirement makes large signal yields unattractive. Hence, a most probable
signal height of about 15 photon equivalents (p.e.) is considered optimal
\cite{CALICE:AHCAL}. By an appropriate choice of the bias voltage of the photon
sensor a signal yield in this region was achieved in the CALICE AHCAL prototype
using tiles with embedded wavelength shifting fiber \cite{CALICE:AHCAL}.

To quantify the non-uniformity, we defined a range of $\pm5\%$, $\pm10\%$ and
$\pm20\%$ around the {\it OMSH} and determined the fraction of the tile area with
{\it MSH} values within the respective ranges. The quality of the tile uniformity
is judged by these area fractions, with the goal to achieve a large area of the
tile within the $\pm5\%$ band. Note, however, that the values for all ranges will
always be below $100\%$ when taking the full tile area into
account because of the finite size of the collimator opening of the radioactive
source and the geometrical acceptance of the trigger scintillator,
which leads to the inclusion of electrons which only penetrate part of the tile in
the edge regions in the signal sample. This results in smeared measurement at the tile edges.
To eliminate those setup specific edge effects, we provide
a second set of area fractions in which measurements at the tile edges were cut,
and a region of $ 1.25\,\text{mm} $ around the tile edges was ignored. In this
case, the effectively investigated tile area is reduced to $
27.5\times27.5\,\text{mm}^2 $.

\section{Optimization of the Tile Geometry}

Scintillator cells with direct fiberless SiPM coupling are only applicable in
highly granular calorimeters if their architecture allows dense packing of the
cells to eliminate inactive zones, if they deliver a single particle signal
amplitude in the desired range providing a good signal-to-noise ratio and a high
dynamic range, and if they achieve a high uniformity of the signal response with
respect to the position at which a particle traverses the tile. These
requirements can be fulfilled by modifications of the tile geometry. The
machining procedure has to be kept simple to allow for mass production of the
tiles.

A gradual reduction of the scintillating material close to the photon sensor
should reduce the overshoot of the response observed for unmodified tiles of the
type ST5. Such a concept was already successfully applied in tile optimization
studies with bottom face coupling \cite{Blazey:2009zz}. To provide full
compatibility with the layer-integrated electronics board for the next generation
HCAL prototype currently under development at DESY \cite{CALICE:Electronics}
within the CALICE collaboration, we adapted this approach to the coupling of the
photon sensor to the center of one side face. To allow seamless packing of the
scintillator cells, the SiPM was embedded into the plastic material.

\begin{figure}
	 \centering
	 \includegraphics[width=0.6\linewidth]{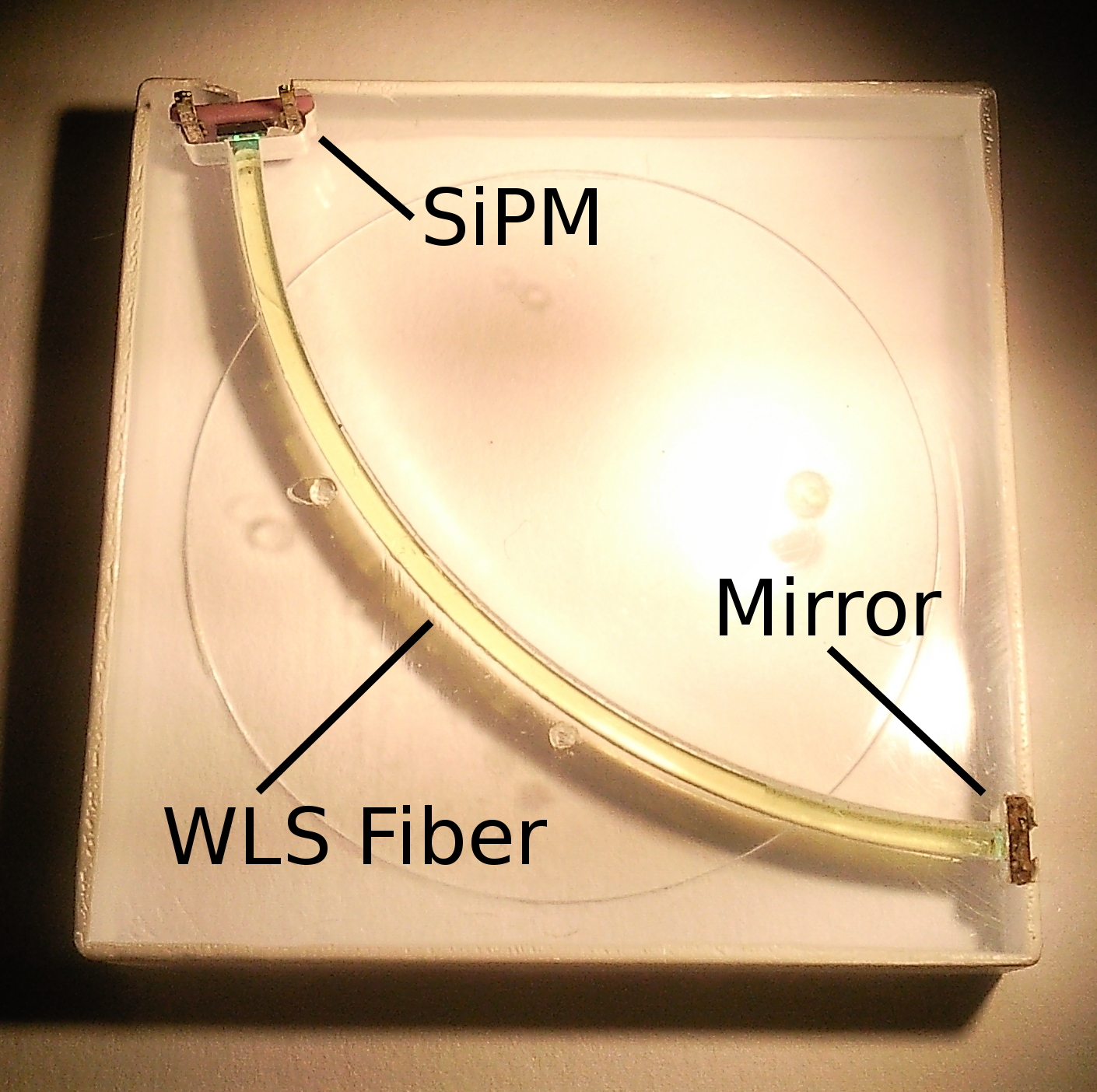}
	 \caption{Photograph of a CALICE 1st generation tile with
	 embedded wavelength shifting fiber which shifts the scintillation light from the blue to the
	 green part of the visible spectrum and leads it to the photosensor.}
	 \label{fig:Analysis:WLSTile}
\end{figure}

\begin{figure}
	  \centering
	  \includegraphics[width=.9\linewidth]{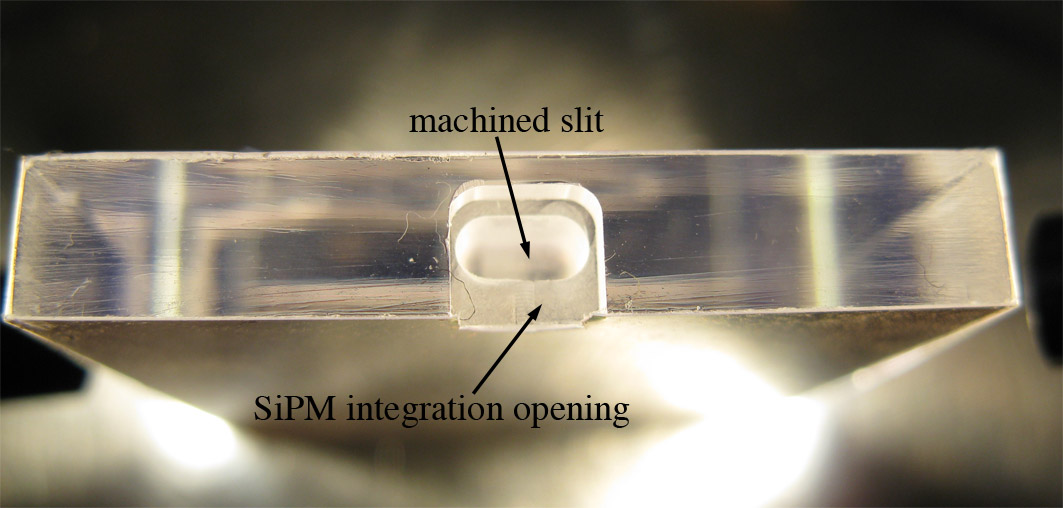}
	  \caption{Photograph of a modified 5 mm thick scintillator tile, 
	  with the integration hole for the photon sensor and the slit 
	  machined into the scintillator material to improve the uniformity 
	  of the response.}
	  \label{fig:ModifiedTile}
\end{figure}

\begin{figure}
	  \centering
	  \includegraphics[width=.6\linewidth]{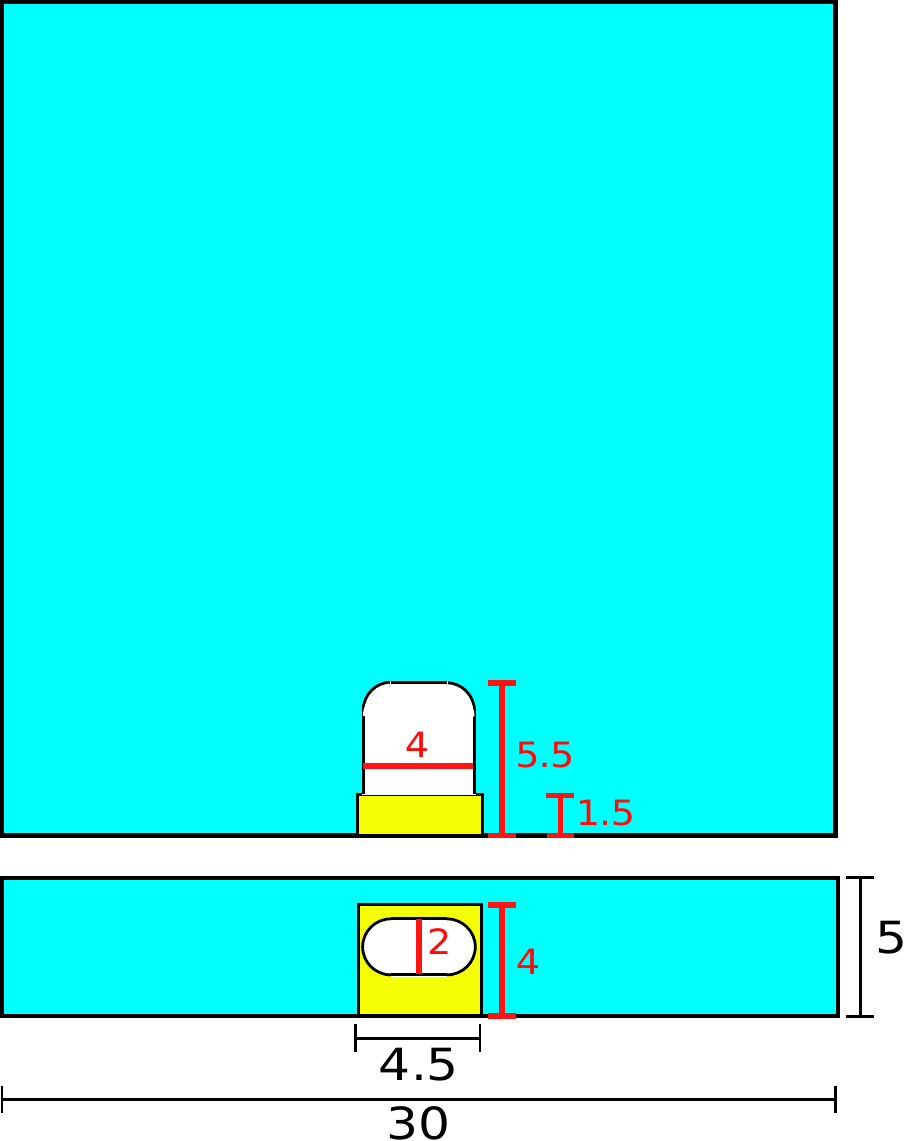}
	  \caption{Dimensions (in mm) of the 5 mm thick scintillator tile with a 
	  geometry optimized for response uniformity and signal amplitude, as 
	  discussed in the text. A 5.5\,mm deep, 4\,mm wide and 2\,mm high slit 
	  to improve the uniformity and a 1.5\,mm deep hole with a cross section 
	  of $4.5 \, \times \, 4 \, \text{mm}^2$ to allow the full integration of 
	  the photon sensor were drilled into the scintillator material.}
	  \label{fig:Tilemod:DimpleDim}
\end{figure}

In several iterations with varying geometries, an optimized tile design was
developed. Figure \ref{fig:Tilemod:DimpleDim} illustrates the overall concept of
the tile design that was further investigated here. A slit with the dimensions of
$2\,\text{mm}$, $4\,\text{mm} $ and $5.5\,\text{mm} $ in height, width and depth
was drilled into the tile. Figure \ref{fig:ModifiedTile} shows the modified tile,
including the integration opening for the photon sensor and the machined slit.
The slit has to reach significantly into the tile to deflect an increased
fraction of propagating photons into the direction of the SiPM. Additionally, the
drilling procedure leaves a fine surface structure on the dimple which refracts
incident light diffusely and disperses it likewise for near and distant particle
penetration positions. This slightly increases the collected light signal per
particle while reducing non-uniformities.

\begin{figure*}
	 \centering
	 \includegraphics[width=0.85\linewidth]{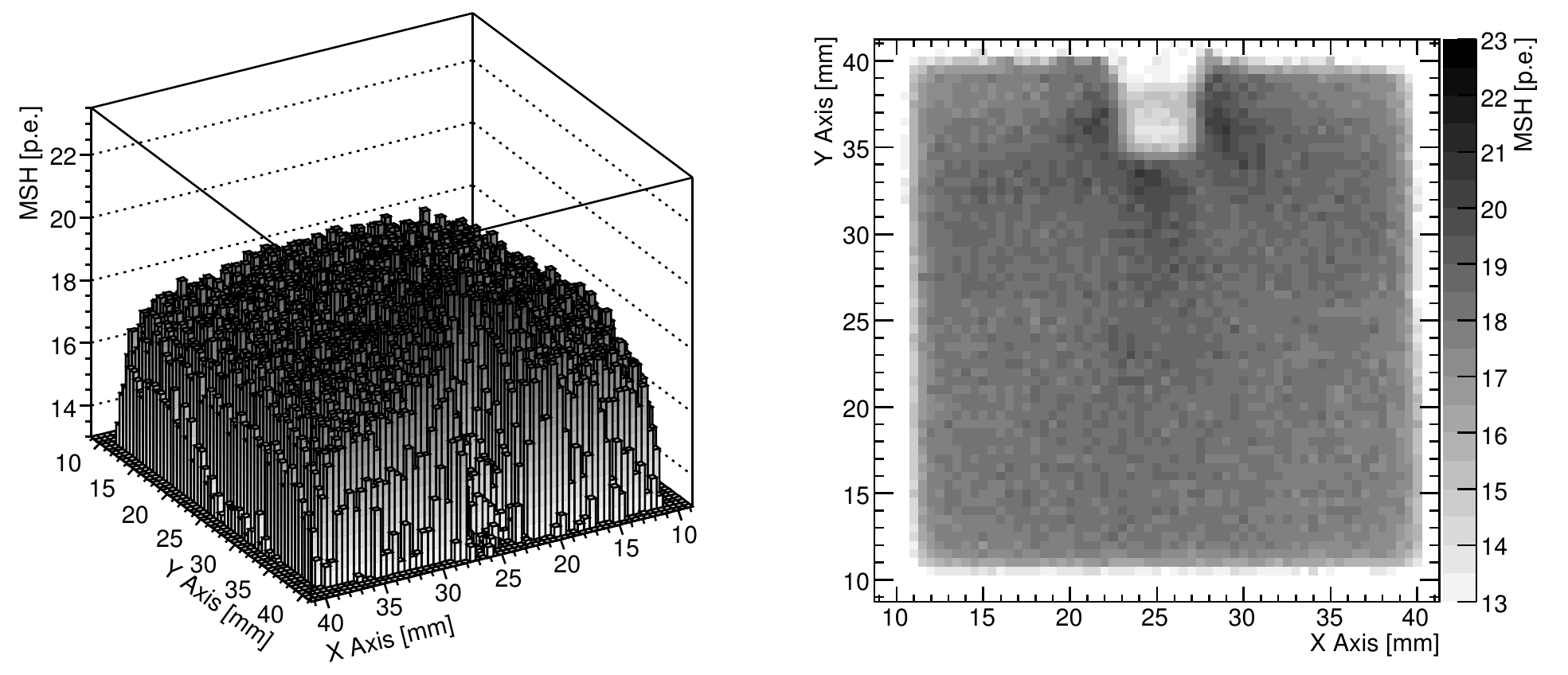}
	 \caption{Modified tile with drilled slit and integrated photon sensor: 
	 Mean signal height of minimum ionizing electrons in dependence of the 
	 penetration position (2D and 3D view), demonstrating a high degree of 
	 uniformity over the full active area.}
	 \label{fig:Analysis:ModTile}
\end{figure*}

To allow close packing of the tiles, as required for the active layers in a
calorimeter, the photon sensor has to be integrated into the scintillator cell to
provide a flat tile face. This is achieved by machining a  rectangular hole that
matches the dimensions of the SiPM casing into the tile side face at the position
of the slit, as shown in Figure \ref{fig:Tilemod:DimpleDim}. This hole also
provides the alignment of the sensitive surface of the SiPM to the center of the
machined slit. After the insertion of the SiPM, the side face of the tile was
completely covered with reflective foil. This integration of the photon sensor
increased the light collection efficiency significantly compared to the external
coupling used for the ST5 tile. In contrast to a tile with embedded WLS 
fiber, as used in the present CALICE physics prototype
\cite{CALICE:AHCAL} (see Figure \ref{fig:Analysis:WLSTile}),
 where the SiPM has to be aligned precisely with the fiber to ensure a full coverage of the fiber cross
section, the tolerances on the SiPM positioning are very generous in case of our
presented tile design. The performance of the SiPM-tile entity remains stable,
provided the SiPM `looks' through the drilled slit into the tile. Tiles with the
modifications discussed here will be referred to in the following as `tile with
slit and SiPM integration' or, in short, 'SI'. Both 5 mm thick tiles (SI5) and 3
mm thick tiles (SI3) have been studied extensively.

\section{Results}

The performance of scintillator tiles with the modifications discussed above and
with a thickness of $ 5\,\text{mm} $ (SI5) and $ 3\,\text{mm} $ (SI3) has been
quantified and compared to the performance of unmodified tiles of the type ST5
with respect to the {\it OMSH} and the uniformity. Table
\ref{tab:Tilmod:TileGeom:SFC} summarizes the measurements of the overall mean
signal amplitude and of the uniformity of the response over the tile surface.

The value of the {\it OMSH} of the modified SI5 tile increased by $42\%$ compared
to the  ST5 tile from $ 13.0\,\text{p.e.} $ to $18.4\,\text{p.e.} $, reaching the
optimal range for excellent signal and noise separation. Also the uniformity of
the tile response was significantly improved by the modifications, as shown in
Figure \ref{fig:Analysis:ModTile}. The response to penetrating particles was
constant to a good approximation over most of the tile area, apart from a minimal
signal overshoot at positions in direct proximity of the slit machined into the
scintillator and a decrease in the area of the
slit itself which amounts to $25\%$ of the {\it OMSH}. As expected, the response is
decreased at the position where the SiPM was integrated into the tile, but note
that it drops nowhere below a value of $ 12.5\,\text{p.e.}$, demonstrating that
the whole tile area was highly sensitive to traversing particles. Comparing the
values for the area fractions listed in Table \ref{tab:Tilmod:TileGeom:SFC}, we
find that the uniformity of a SI5 tile improves by $27\%$ and $4\%$ in the $\pm5\%$ and $\pm10\%$ region,
respectively.

\begin{table}
	\centering
	\begin{tabularx}{\linewidth}{|X|X|X|X|X|X|} \hline
		Tile & {\it OMSH} & $\pm$5\,\% & $\pm$10\,\% & $\pm$20\,\%\\ \hline \hline
 		ST5 & 13.0\,p.e. & 57.4\,\% & 80.8\,\% & 90.7\,\% \\	\hline
 		ST5 (cut) & & 69.1\,\% & 93.8\,\% & 98.3\,\% \\	\hline
 		SI5 & 18.4\,p.e. & 72.7\,\% & 84.2\,\% & 90.1\,\% \\	\hline
 		SI5 (cut) & & 87.9\,\% & 97.1\,\% & 98.9\,\% \\	\hline
 		SI3 & 13.0\,p.e. & 70.3\,\% & 84.3\,\% & 94.0\,\% \\	\hline
 		SI3 (cut) & & 82.5\,\% & 94.0\,\% & 98.0\,\% \\	\hline
	\end{tabularx}
	\caption{The uniformity of three different tiles: Unmodified tile with
	direct SiPM coupling (ST5) and tiles with applied slit and integrated SiPM
	coupling in an $ 5\,\text{mm} $ (SI5) and $ 3\,\text{mm} $ (SI3) option.
	Shown is the overall mean of the signal height ({\it OMSH}) and the tile area within
	a deviation from this value in three different ranges. The values
	labelled ``cut'' are determined over a reduced area of the tile, excluding the edge
	regions which suffer from inaccuracies due to the acceptance of the setup, as discussed in the text.}
	\label{tab:Tilmod:TileGeom:SFC}
\end{table}

The same modifications were also investigated for tiles of $ 3\,\text{mm} $
thickness (SI3), and show a comparable performance in terms of response
uniformity, as summarized in Table \ref{tab:Tilmod:TileGeom:SFC}. Due to the
reduced thickness, the overall signal amplitude was reduced compared to SI5, and
was in the same range as for the unmodified directly coupled 5 mm thick tile ST5.
Thinner tiles are of considerable interest for calorimeters at future collider
detectors, since they allow a higher average density of the detector for a given
sampling frequency. This makes more compact calorimeters possible, reducing the
required radius of the experiment's solenoidal magnet for a given maximum energy
leakage, resulting in significant cost savings.

\section{Conclusion}

Direct fiberless coupling of silicon photomultipliers to plastic scintillator
tiles, made possible by blue sensitive SiPMs, allows significant simplifications
of the scintillator tile construction, and consequently potential cost and time
savings in the construction of highly granular analog calorimeters. However,
specific modifications of the tile geometry are necessary to provide a large
signal for fully penetrating minimum-ionizing particles and a high degree of
uniformity of the response over the full active area.

We have developed specific modifications for tiles with SiPMs coupled to the
center of a side face of the cell. These consist of a drilled slit in the
scintillator and a hole to allow the full integration of the photon sensor, to
achieve a high degree of uniformity and a satisfactory signal amplitude for both
5 mm thick and 3 mm thick scintillator tiles with lateral dimensions of $3 \,
\times \, 3\,\text{cm}^2$. The direct coupling also significantly relaxes the
required alignment precision of the photon sensor, simplifying the assembly and
the mass production of scintillator tile - SiPM units. The studies were performed
in the context of highly granular hadron calorimetry, but the developed
scintillator cell design is also suitable for other applications in which a high
level of integrability, cell uniformity or efficiency is needed for particle
detection.


\end{document}